\documentclass[aps,prl,reprint,superscriptaddress,showpacs,showkeys,longbibliography]{revtex4-1}
\usepackage{amssymb}
\usepackage{amsmath}
\usepackage{cancel}
\usepackage{color}
\usepackage{braket}
\usepackage{graphicx}
\usepackage{epstopdf}
\usepackage{verbatim}
\usepackage{siunitx}
\usepackage{tikz}
\usepackage{soul}
\usepackage{xr}
\usepackage[toc,page]{appendix}
\usepackage[section]{placeins}
\usepackage[T1]{fontenc}
\usepackage{currvita}

\makeatletter
\def\clearfmfn{\let\@FMN@list\@empty}    
\makeatother

\renewcommand{\check}[1]{\textcolor{red}{#1}}

\newcommand{\harvard}{Department of Physics, Harvard University, 17 Oxford Street, Cambridge, Massachusetts 02138, USA}

\begin{document}

\title{
Strong coupling of two individually controlled atoms via a nanophotonic cavity
}
\author{Polnop Samutpraphoot}
\thanks{These authors contributed equally to this work}
\affiliation{\harvard}
\author{Tamara \DJ or\dj evi\'c}
\thanks{These authors contributed equally to this work}
\affiliation{\harvard}
\author{Paloma~L.~Ocola}
\thanks{These authors contributed equally to this work}
\affiliation{\harvard}
\author{Hannes~Bernien}
\affiliation{Pritzker School of Molecular Engineering, University of Chicago, Chicago, IL 60637, USA}
\author{Crystal~Senko}
\affiliation{Department of Physics and Astronomy, University of Waterloo, Waterloo, N2L 3R1, Canada}
\affiliation{Institute for Quantum Computing, University of Waterloo, Waterloo, N2L 3R1, Canada}
\author{Vladan Vuleti\'{c}}
\affiliation{Department of Physics and Research Laboratory of Electronics,
Massachusetts Institute of Technology, Cambridge, MA 02139, USA}
\author{Mikhail D. Lukin}
\email[]{lukin@physics.harvard.edu}
\affiliation{\harvard}

\begin{abstract}
We demonstrate photon-mediated interactions between two individually trapped atoms coupled to a nanophotonic cavity. Specifically, we observe superradiant line broadening when the atoms are resonant with the cavity, and level repulsion when the cavity is coupled to the atoms in the dispersive regime.
Our approach makes use of individual control over the internal states of the atoms,
their position with respect to the cavity mode,
as well as the light shifts to tune atomic transitions individually,
allowing us to directly observe the anti-crossing of the superradiant and subradiant two-atom states.
These observations open the door for realizing quantum networks
and studying quantum many-body physics based on atom arrays coupled to nanophotonic devices.
\end{abstract}

\maketitle


Controlled interactions between individual photons and quantum emitters
are an important ingredient for the realization of scalable quantum information systems
\cite{divincenzo2000criteria,
reiserer2015review}.
Nanophotonic devices in which the light is confined to sub-wavelength
dimensions constitute a promising approach for engineering strong light-matter coupling
\cite{lodahl2015review,
chang2018review}.
The appeal of a nanophotonic platform is due to two key features. On one hand,
nanophotonic devices allow the photonic dispersion to be tailored
to achieve tunable-range interactions between coupled emitters
\cite{
lodahl2004quantumdotpc,
englund2010quantumdotpc,
thompson2013couplingpc_nourl,
sipahigil2016sivnanophotonics_nourl,
miyazono_faraon2016erbiumnanophotonics,
dibos_thompson2018erbiumnanphotonics}
and engineer a range
of interaction Hamiltonians
\cite{doulgas2015bandtuning,
gonzaleztudela2015simulation}.
On the other hand, the nanoscale mode volume
enables an efficient high cooperativity emitter-photon interface
suitable for realizing potentially scalable systems
\cite{duan2004photongate,
duan2003atomentanglement,
briegel_zoller1998repeater,
duan2001dlcz}.
In particular, cold neutral atoms have recently emerged as a promising approach
for realizing large-scale quantum systems due to the ability to generate large numbers of identical,
individually trapped atoms
\cite{kim_anh2016array,
endres2016array_nourl,
kim_anh2016array,
barredo_browaeys2016array2d_nourl,
barredo_browaeys2018array3d,
kumar_weiss2018maxwelldemon}.
While significant effort is currently being directed towards coupling multiple
isolated atoms to nanophotonic systems
\cite{goban2015superradiance,
kim_clhung2019chiptrapping,
reitz_rauschenbeutel2010fibercoupling,
thompson2013couplingpc_nourl},
achieving a strong coupling of a deterministic number of atoms remains a challenge.
The atoms must be trapped closely enough to the device
to maximize the coupling within the evanescent field, while
overcoming attractive surface forces
\cite{antezza_stringari2004casimirtheory,
obrecht_cornell2007casimir},
and preserving the excellent atomic coherence properties.

\begin{figure}[h!]
\includegraphics[width=9cm]{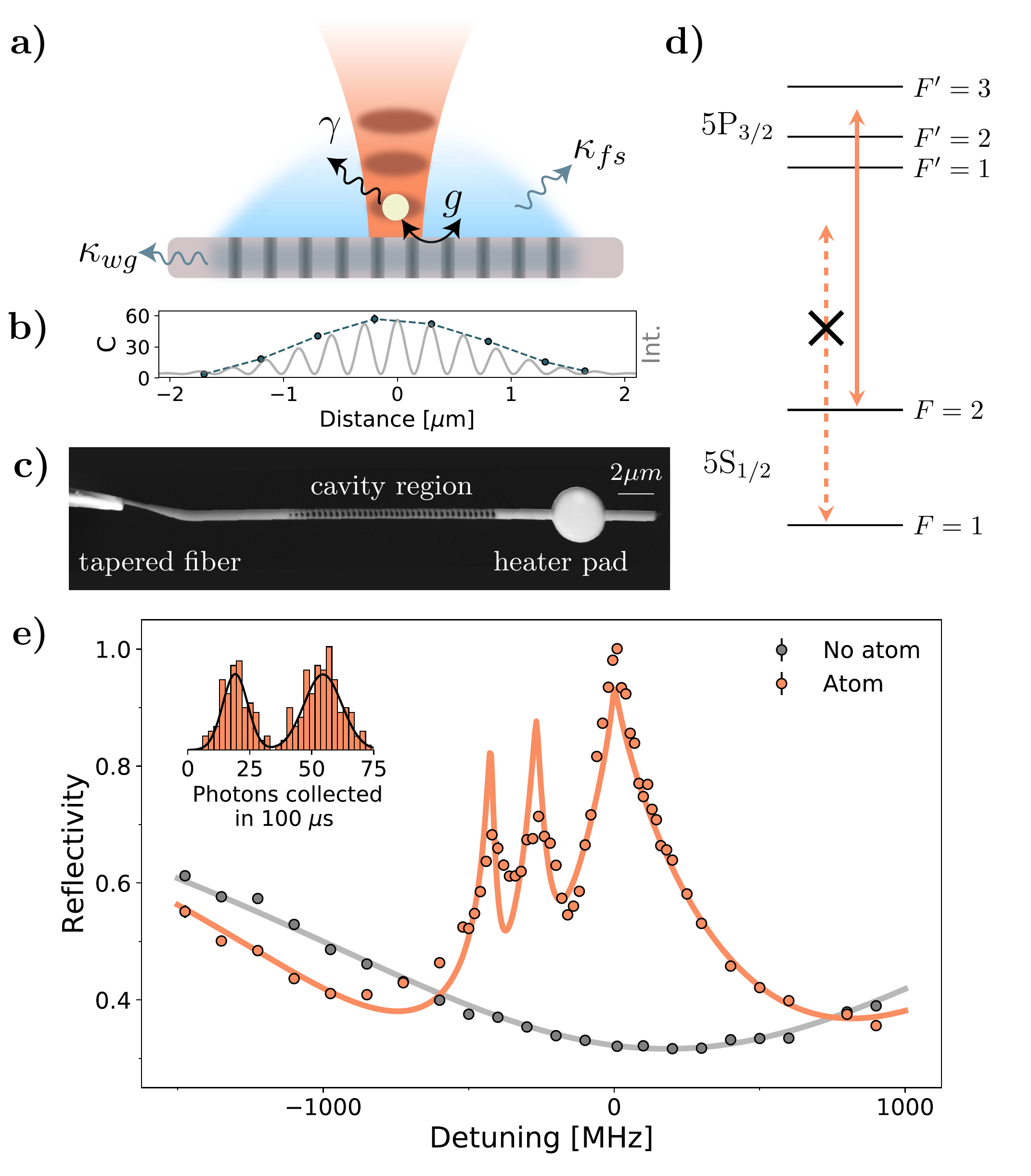}
\vspace{-15pt}
\caption{ {\bf High cooperativity atom-photon coupling to a nanophotonic cavity.}
{\bf a)} Schematic of the experimental setup,
showing an atom trapped in the lattice of an optical tweezer
coupled to a nanophotonic cavity. The parameters are defined in the main text.
{\bf b)}
Moving the tweezer along the cavity to map out the mode in terms of cooperativity (blue)
and simulated intensity profile of the cavity mode (gray).
{\bf c)} SEM image of the nanophotonic cavity suspended on a tapered fiber.
{\bf d)} Level diagram for the $5S_{1/2} \rightarrow 5P_{3/2}$ transition.
The $F= \{2, 1 \}$ manifold is \{coupled, uncoupled\} to the cavity.
The probe detuning is defined relative to the bare $F=2 \rightarrow F' =3$ transition.
{\bf e)} Measured cavity reflection spectrum with and without an atom
coupled to the cavity.
Inset: histogram of counts collected at 0~MHz detuning, showing single-shot atomic detection.
\label{fig1}}
\vspace{-10pt}
\end{figure}

In this Letter, we report on the observation of strong coupling of two
individually controlled atoms via a nanophotonic cavity.
In particular, we spectroscopically demonstrate
superradiant line broadening in the resonant regime,
and level repulsion in the dispersive regime.
These experiments utilize individual control of
the positions of the atoms with respect to the cavity mode,
 their internal states, and the frequencies of
their transitions. This allows us to observe the anti-crossing
of the superradiant and subradiant two-atom states, in analogy to prior
observations involving superconducting qubits and color centers in diamond
\cite{majer_schoelkopft2007superconductinginteraction,
evans2018avoidedcrossing_nourl}.

\begin{figure}
\includegraphics[width=9.3cm]{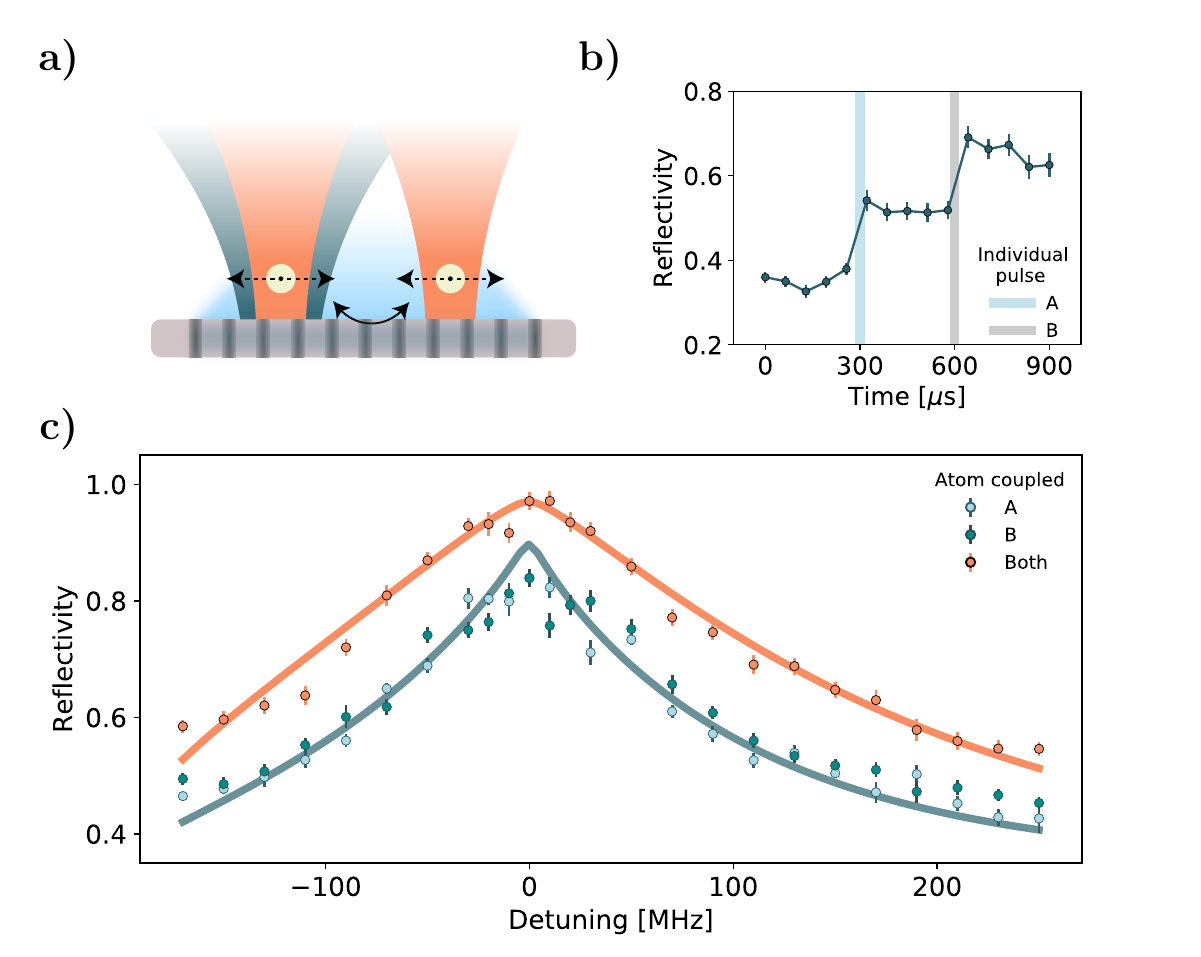}
\vspace{-15pt}
\caption{ {\bf Individual control and resonant coupling.}
{\bf a)} Illustration of the individual position and internal state control with a repumping (blue) beam co-propagating with one of the optical tweezers (orange).
{\bf b)} Average reflectivity jumps after pumping each atom to the $F=2$ manifold,
taken at the probe detuning of $100~\text{MHz}$.
{\bf c)} Superradiant linewidth enhancement of the
$F=2 \rightarrow F' =3$ line.
The labels \{A, B\} denote the atoms in the two tweezers.
The spectrum is taken when the tweezers are positioned $1~\mu\text{m}$ away
from the mode center to retain addressability.
\label{fig2}}
\end{figure}


Our experiments utilize a cavity QED system
consisting of $^{87}$Rb atoms coupled to the
evanescent field of a photonic crystal (PC) cavity
(Fig.~\ref{fig1}a, b) \cite{thompson2013couplingpc_nourl}.
The SiN nanophotonic device is suspended in a vacuum chamber
on a tapered optical fiber, which is also used for efficient
interrogation of the cavity by exciting
and collecting photons through its fiber-optic interface
\cite{tiecke2015taperedfiber}.
This approach minimizes the physical footprint of the system,
thus allowing for good optical access and unobstructed
trapping and cooling while retaining the flexibility
to control the atomic position and confinement.
The atoms are trapped using tightly focused optical
tweezers that localize them to within tens of nanometers
and can be steered to desired locations.
The internal states of the atoms are individually
manipulated with light fields co-propagating with the optical tweezers.
We interrogate the response of the atom-cavity system
by scanning the frequency of the probe field to measure the reflection spectrum.
The frequency of the cavity is tuned
thermally with a laser beam pointed at the heater pad (Fig.~\ref{fig1}c).


We probe the atom-cavity spectrum by tuning the cavity
to the $5S_{1/2} \rightarrow 5P_{3/2}$ transition at 780 nm.
The spectrum is acquired by scanning the probe field from the ground
state manifold $5S_{1/2},~F=2$
across the excited state manifolds $5P_{3/2},~F'=1,2,3$ (Fig.~\ref{fig1}d),
while the atom is positioned at the center of the cavity mode.
We note that for all the spectra presented in this Letter,
the probe detuning is relative
to the bare $F=2 \rightarrow F' =3$ transition,
and the error bars are obtained from statistical uncertainties
acquired over multiple experimental runs.

\begin{figure}[htb]
\includegraphics[width=8.5cm]{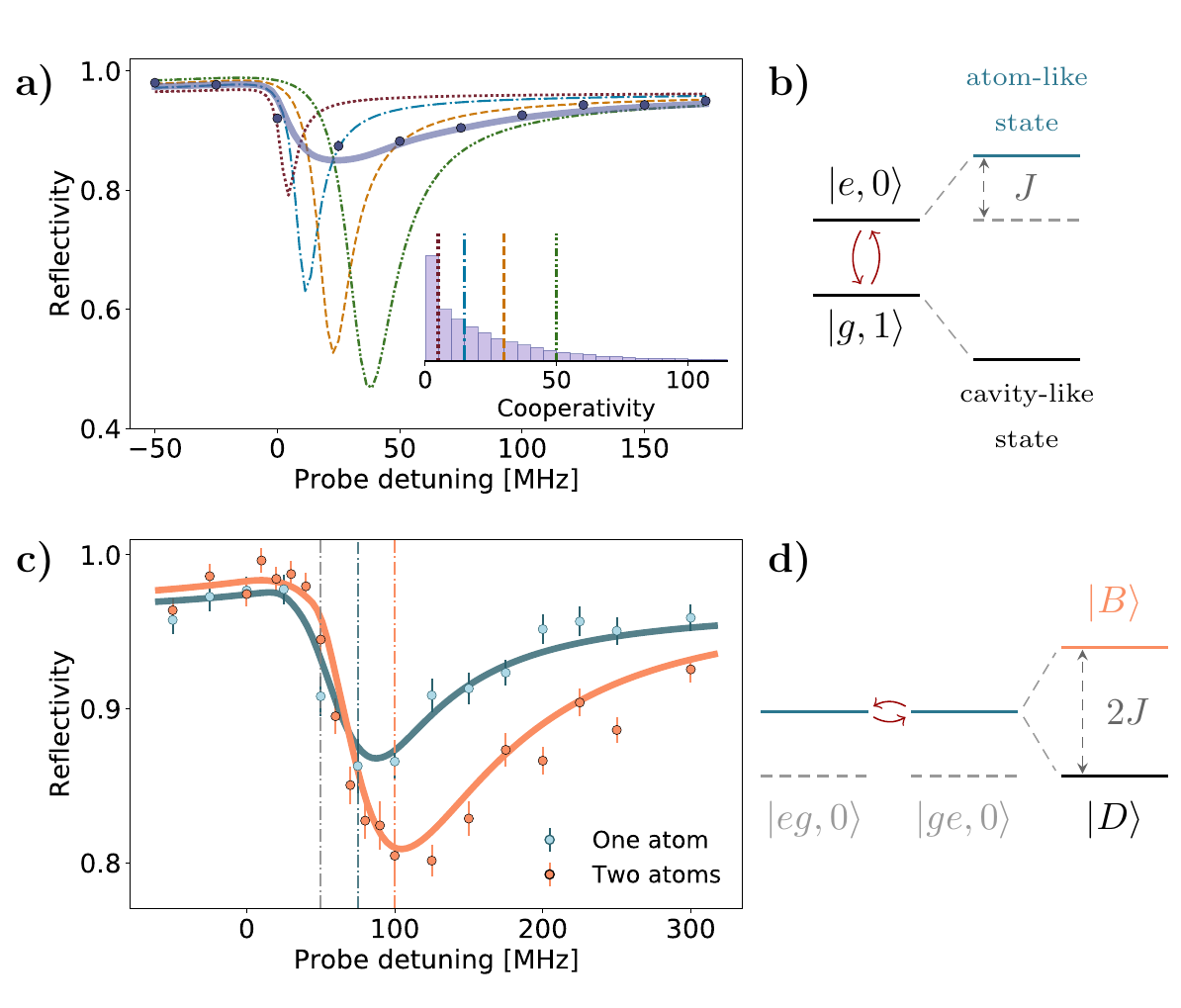}
\vspace{-0pt}
\caption{{\bf Atom-photon interaction in the dispersive regime.}
{\bf a)}
Atom-like spectrum of the $F=2\rightarrow F'=3$ line at cavity detuning
$\Delta=2\kappa$.
Theoretical curves (dashed-dotted) generated with
single-valued cooperativities (vertical lines, inset).
Theoretical curve (solid) constructed using cooperativity distribution (inset).
This spectrum is acquired without a light shift from the tweezer.
{\bf b)} Level diagram of one atom in the singly-excited manifold.
Bare states with \{atomic, photonic\} excitation $\{\ket{e,0},~\ket{g,1}\}$
are dressed by the atom-photon coupling $J = g^2 /\Delta$
into the \{atom,~cavity\}-like components.
{\bf c)}
Single- and two-atom spectra when the two atoms are resonant with each other.
The vertical lines delineate the frequency of an atom in a
$50~\text{MHz}$-deep tweezer not coupled to the cavity (gray)
and the expected cavity-induced shifts of the
\{one, two\} atom(s) coupled to the cavity \{blue, orange\}.
{\bf d)}
Level diagram of two atoms. The two atom-like components
hybridize into the \{dark, bright\} states \{$\ket D$, $\ket B$\}
which experience frequency shifts of \{0, $2J$\}.
\label{fig3}}
\end{figure}

Fig.~\ref{fig1}e shows the reflection spectrum of the cavity
with and without an atom present.
For the case without an atom, the reflection
spectrum is a resonance dip of the empty cavity.
The presence of an atom drastically changes the spectrum,
and we observe three atomic lines that are significantly
broadened due to resonant coupling between the atom and the cavity.
This effect can be understood
in terms of the Purcell enhancement.
In the resonant regime, the radiative
decay rate into the cavity mode
is enhanced by the single-atom cooperativity $C = 4g^2/\kappa\gamma$,
where $g$ is the single-photon Rabi frequency, $\gamma$
the atomic spontaneous decay rate. The cavity decays at the rate
$\kappa_{wg}$ into the waveguide and $\kappa_{sc}$ elsewhere,
yielding the total cavity decay rate
$\kappa = \kappa_{wg} + \kappa_{sc}$ (Fig.~\ref{fig1}a).
The observed lineshape is accurately described using a model incorporating
a distribution of cooperativities resulting from
fluctuations in the atomic positions
(see Fig.~\ref{fig3} and \cite{si} for further discussions).
Taking these considerations into account,
we fit the spectrum in Fig.~\ref{fig1}e and extract the average cooperativity
\mbox{$C=71(4)$}, corresponding to the cavity QED parameters
\mbox{$\{2g,\gamma,\kappa_{wg},\kappa_{sc}\}
=2\pi\times\{1.24(4),0.006,0.86,2.77\}~\text{GHz}$}.
Given the geometry of the trapping potential of the tweezer and the evanescent field
of the cavity, we estimate that the atom in the closest lattice site at a distance of
260~nm from the surface of the PC experiences a single-photon Rabi
frequency $2g_0 = 2\pi\times1.7$~GHz in the absence of the cooperativity fluctuation,
corresponding to $2g = 2\pi\times1.26$~GHz after averaging \cite{si}.
This estimate is consistent with the single-photon Rabi frequency
extracted from our experimental data.


The efficient atom-photon interface allows us to determine the presence of an atom in a single shot.
We tune the probe frequency to the $F=2 \rightarrow F'=3$ line
and count reflected photons collected within $100 ~ \mu\text{s}$
(Fig.~\ref{fig1}e inset).
When repeated multiple times, the photon number follows a bimodal normal distribution
with $0.7\%$ overlap, which is adequately separable to determine if
the atom is coupled to the cavity.
The atom becomes uncoupled from the cavity if
it falls into the $F=1$ manifold via off-resonant scattering.
We deplete the population in the $F=1$ manifold by sending in an additional beam co-propagating with the optical tweezer on the
\mbox{$5S_{1/2},~ F = 1 \rightarrow 5P_{1/2},~ F' = 2$} transition,
at 795 nm, which is sufficiently detuned from the cavity and can be
filtered out from the collected photons.
In addition to the $F = 2 \rightarrow F'=3$ line,
the spectrum in Fig.~\ref{fig1}e
also shows the $F = 2 \rightarrow F'=1,~2$ transitions, which are not cycling
and would not be visible without applying the repumping beam.


The reflection spectrum can be used to study the cooperativity dependence
on experimental parameters such as the position of the atom.
Taking advantage of the individual position control,
we scan the tweezer position along the
axis of the PC by steering the galvanometer mirrors and
acquire a spectrum associated with each position
\cite{si}.
We then determine the cooperativity at each location from the
Purcell-enhanced linewidth
and find that the resulting cooperativity
dependence (Fig.~\ref{fig1}b, blue) traces out
the envelope of the numerically simulated field profile
intensity (gray).

Having characterized the single-atom coupling,
we now turn to the case of two atoms
to study their resonant coupling via the cavity
(Fig.~\ref{fig2}a). The two tweezers must be placed
away from each other to avoid overlap and cross-talk,
hence inevitably lowering their individual cooperativities.
The experiments involving two atoms
are operated with the tweezers placed
$1~\mu\text{m}$ away from the mode center, corresponding
to the average single-atom cooperativity
$C=31(2)$ (Fig.~\ref{fig1}b, blue),
well within the strong coupling regime.

In addition to the position control,
the two-atom experiments also make use of
internal state manipulation of individual atoms.
This is achieved by having repumping beams co-propagating with the tweezers,
selectively bringing the desired atom into the $F=2$ manifold.
We demonstrate this by applying $3~\mu\text{s}$ long
repumping pulses at $300~\mu\text{s}$ on one tweezer
and $600~\mu\text{s}$ on the other while constantly probing the spectrum with the
probe blue-detuned to 100~MHz and monitoring the collected photon counts in
time (Fig.~\ref{fig2}b).
The average collected counts step up following each pulse,
indicating the influence of repumping the atoms one at a time.
The capability of individual repumping is utilized for detection and postselection on having two atoms coupled to the cavity in each trial of the experiment \cite{si}.

With these capabilities, we explore the collective behavior of two atoms simultaneously coupled to the cavity by acquiring a reflection spectrum centered on the $F~=2\rightarrow F'=3$ line (Fig.~\ref{fig2}c, orange).
As compared to the single-atom case, we observe twofold broadening of the two-atom line, which can be interpreted as a result of superradiance
\cite{goban2015superradiance,
wallraff2014superradiance}.
The spectrum is consistent with our theoretical model generated using the sum of individual cooperativities extracted from the single-atom spectra with no additional parameters \cite{si}.

\begin{figure*}[htb]
\includegraphics[width=18.0cm]{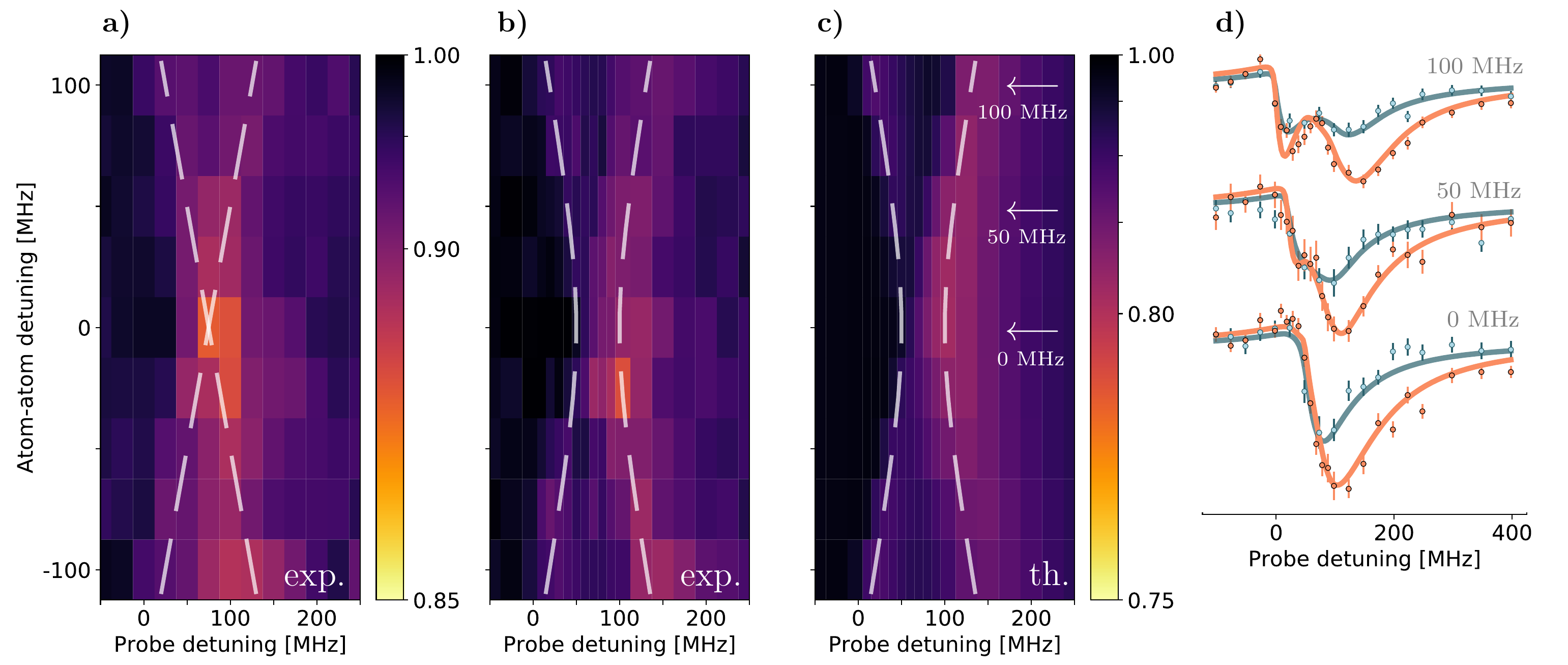}
\vspace{-15pt}
\caption{ {\bf Level repulsion of the atomic lines induced by the cavity-mediated interaction.}
{\bf a)} Experimental data of reflectivity
map over probe detuning and relative atom detuning $\delta_{AB}$,
taken by averaging two independent spectra and
plotted with the atoms' frequencies (dashed line), showing level crossing.
{\bf b)} Experimental data for two atoms simultaneously coupled to the cavity,
plotted with frequencies of the bright and dark states (dashed line), showing level repulsion.
{\bf c)} Theoretical calculation for two atoms simultaneously coupled to the cavity.
{\bf d)} Cross-sections of
the spectra at $\delta_{AB} = 2\pi\times \{0, 50, 100\}~\text{MHz}$
indicated by the white arrows in Fig. 4c.
Average of single atom spectra (blue) and two atom spectra (orange).
\label{fig4}}
\end{figure*}

The phenomena described above, such as the Purcell effect
and superradiant broadening, can also be demonstrated spectroscopically
in the dispersive regime.
We red-detune the cavity relative to the $F=2 \rightarrow F'=3$
line by $\Delta = 2\kappa$ and acquire a single-atom spectrum (Fig.~\ref{fig3}a).
We observe that the atomic line becomes narrower than the resonant case
and experiences a frequency shift
\cite{0mhzshift}.
The shift can be interpreted
as a result of the atom-cavity interaction
in the dispersive regime (Fig.~\ref{fig3}b).
In this regime, the atom is dressed by the cavity coupling
and experiences a frequency shift of
$g^2/\Delta = C \kappa \gamma /4\Delta$ while its Purcell-enhanced linewidth is
suppressed by a factor of $1+4\Delta^2/\kappa^2$.

We note that similarly to the resonant case, the experimental data cannot be
adequately captured by theoretical curves (Fig.~\ref{fig3}a, dashed-dotted lines)
generated with single-valued cooperativities (Fig.~\ref{fig3}a inset, vertical lines)
but is rather well described using a model involving cooperativity distribution
shown in the inset of Fig.~\ref{fig3}a.
This distribution leads to varying frequency shifts, which result in
additional broadening and modified lineshape.
The origin of the cooperativity distribution
can be understood as fluctuations in the atomic positions
predominantly due to thermal motion across the spatially-varying field of the cavity \cite{si}.

The collective coupling in the dispersive regime can be characterized
by probing spectra at a cavity detuning, here chosen to be $\Delta = 2\kappa$,
with one and two atom(s) at the same resonance frequency (Fig.~\ref{fig3}c).
The single-atom shift is consistent with the expected value of
$J = 2\pi\times25(4)~\text{MHz}$ based on our average cooperativity estimate.
The two-atom spectrum appears as a single line that experiences a shift twice as large as the single-atom line
\cite{50mhzshift}

The twofold enhancement of the frequency shift can be formulated in terms of
level repulsion between the two atoms resonant with each other (Fig.~\ref{fig3}d).
Here, the resulting states can be represented
as symmetric and antisymmetric superpositions of the
two-atom states.
The symmetric superposition interacts more strongly with the cavity
due to constructive interference of coherent scattering into
the cavity mode, thereby experiencing the frequency shift of $2 g^2/\Delta$,
and forming the bright state,
\mbox{$\ket B = (\ket{eg,0} + \ket{ge,0})/\sqrt{2}$}.
The antisymmetric superposition does not interact with the cavity
due to the destructive interference,
acquires zero shift,
and forms the dark state \mbox{$\ket D=(\ket{eg,0}-\ket{ge,0})/\sqrt{2}$}.
The frequency difference hence results in line-splitting
which can be equivalently viewed as an interaction
between two atoms with coupling rate
$2J = 2 g^2/\Delta$ due to an
off-resonant exchange of virtual cavity photons
\cite{goban2015superradiance}.

We further illustrate the dynamics of level repulsion
by tuning the atoms in and out of resonance with each other.
The relative atom detuning
$\delta_{AB} = \delta_A - \delta_B$,
can be tuned with the light shifts induced by the individual
tweezers, in our case, within $ 2\pi\times (\pm100)~\text{MHz}$ \cite{si}.
At large $\delta_{AB}$, the two atoms
remain independent, and the spectrum
shows two separate lines.
If there were no interaction between the two atoms,
the spectra would cross as we tune their resonances past each other
as shown in an average of two individual atomic spectra in Fig.~\ref{fig4}a.
However, when both atoms are coupled to the cavity and are tuned closer together,
the single-atom states start to hybridize into the
dark and bright components that avoid crossing each other (Fig.~\ref{fig4}b).
The frequencies of the states follow
$\sqrt{(2J)^2 + \delta_{AB}^2}$ with the gap that signifies
the coupling strength of $2J = 2\pi\times 50(8)~\text{MHz}$,
in agreement with the theoretical model shown
in Fig.~\ref{fig4}c. In Fig.~\ref{fig4}d,
we show three of the cross-sections of the experimental data
overlapped with our theoretical model, both for the two-atom and the
averaged case.


The observations demonstrate a controllable high-cooperativity interface
between atoms and photons, as well as between two atoms mediated
by virtual cavity photons, namely superradiance and anti-crossing.
These results can be extended along several directions.

First, the photon-mediated interactions can be combined with
coherent quantum control of the internal states of the atoms to
implement quantum gates for state transfer and entanglement generation
\cite{ciraczollermabuchi1997statetransfer,
duankimble2004,
welte2017carving,
welte2018gate}.
As the errors of many of the protocols for these applications
are inversely proportional to the cooperativity, further improvement
in cooperativity is an essential prerequisite for scalability.
This may be achieved by positioning the atoms closer
to the surface to access a larger field strength
\cite{hung2013trappingpc},
or improvements in fabrication and design
\cite{ji_lipson2017ringresonator}.
A cooperativity increase can be achieved with tighter localization with respect
to the mode maxima by cooling the atoms in optical tweezers
\cite{kaufman2012raman,thompson2012raman} and
near the nanostructure
\cite{meng_rauschenbeute2018cooling}.
Second, this approach offers a complete toolbox for
controlling quantum many-body systems.
The number of atoms can be scaled up by generating tweezer arrays
\cite{kim_anh2016array,
endres2016array_nourl,
kim_anh2016array,
barredo_browaeys2016array2d_nourl,
barredo_browaeys2018array3d,
kumar_weiss2018maxwelldemon}.
The established techniques for assembling atom arrays can be combined
with our approach for the individual addressing and light shift control
and recently developed techniques for imaging an array on
a nanophotonic structure \cite{kim_clhung2019chiptrapping}.
Combining these capabilities with the ability to engineer band dispersion
may allow for the exploration of novel many-body systems with extensive tunability.
Finally, the efficient high-bandwidth atom-photon interface
with individual atomic control is naturally suitable for
realizing quantum networks with multi-qubit nodes
\cite{kimble2008quantuminternet,
khabiboulline2018telescope,
komar2014clocks}.
The nanoscale interface also holds prospects of integration
with modular architecture such as on-chip photonic circuits
and fiber-optic networks for various applications ranging from
quantum repeaters to distributed quantum computing
\cite{borregaard2015repeater,
nickerson2013distibutedcomputing}.

We thank
Rivka Bekenstein,
Robert Bettles,
Johannes Borregaard,
Manuel Endres,
Hannes Pichler,
Florentin Reiter,
Sylvain Schwartz,
Fran\c{c}ois Swiadek,
Jeff Thompson,
Susanne Yelin,
and Alexander Zibrov
for useful discussions and experimental contributions. This work
was supported by  the Center for Ultracold Atoms, the National Science Foundation,
AFOSR MURI, Vannevar Bush Faculty Fellowship, and ARL CDQI.


%

\clearpage

\onecolumngrid

\renewcommand{\thefigure}{S\arabic{figure}}
\renewcommand{\thetable}{S\arabic{table}}
\setcounter{figure}{0}
\setcounter{equation}{0}

\begin{center}
  \textbf{\large Supplementary information}\\[.2cm]
  \vspace{0.2cm}
\end{center}

\section{I. Apparatus}

\begin{figure*}[htb]
\vspace{15pt}
\includegraphics[width=17cm]{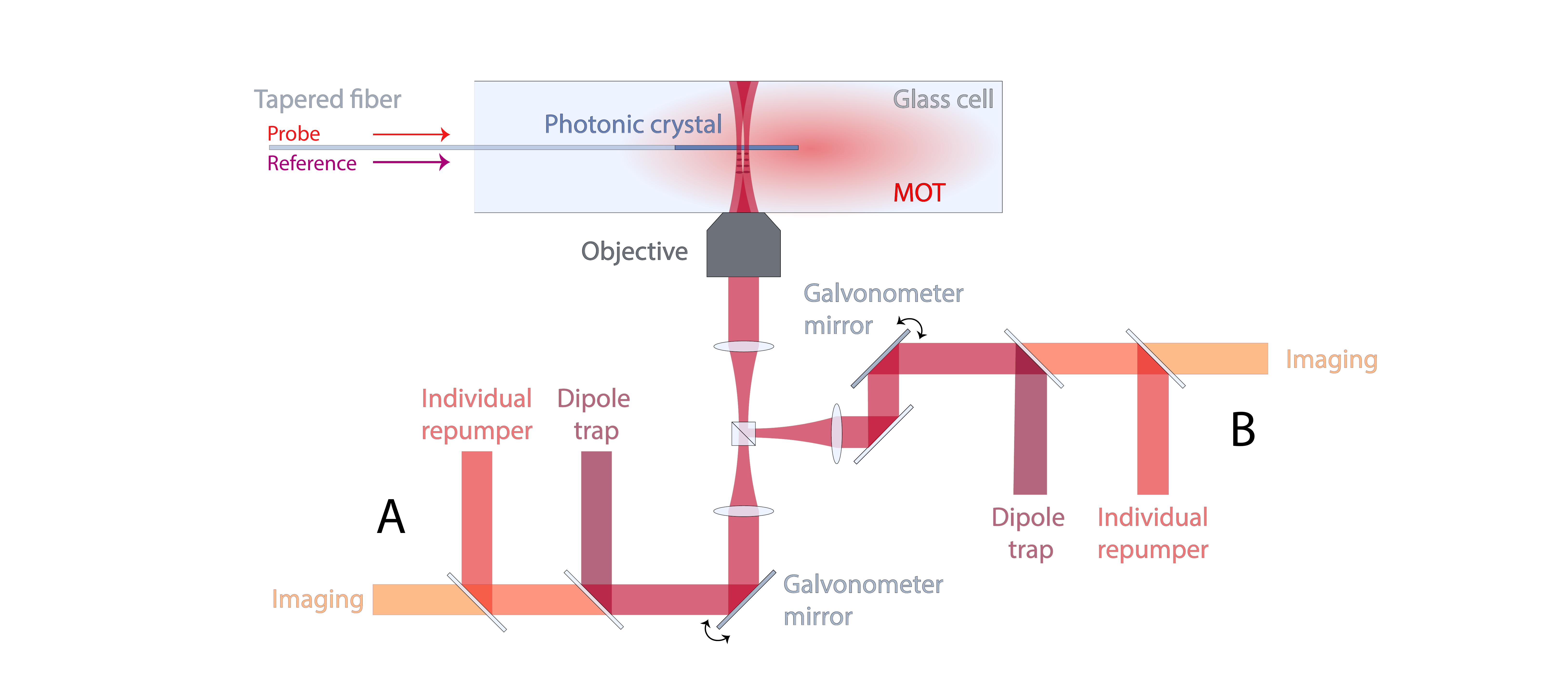}
\caption{Schematic of experimental setup (not to scale) detailing how
the trap paths are combined and focused by the objective onto the PC.
Resonant light is sent through the tapered fiber to the PC to probe the atom-cavity response,
along with far-detuned light for cavity stabilization,
which is continuously monitored in reflection.}
\label{fig_exp_setup}
\end{figure*}

The apparatus consists of a confocal microscope that images a photonic crystal
(PC) installed in the glass cell of the vacuum chamber (Fig. \ref{fig_exp_setup}).
The microscope objective [Mitutoyo G Plan Apo 50X] is used for focusing
optical tweezers (beam waist $w_0\sim$ 900~nm) and for fluorescence imaging of single atoms.
The two tweezers are generated by combining two optical paths on a 50:50
non-polarizing beam splitter at the focus of a $4f$-microscope configuration.
Each path also has two dichroic mirrors, to combine the trap light (815~nm)
with individual repumping light (795~nm) and to separate the collected fluorescence light (780~nm).

The ultra-high vacuum (UHV) system consists of a glass cell mounted to a compact vacuum cube
[Kimball Physics MCF450-SphSq-E2C4] and a combination of a non-evaporable getter
and an ion pump [NEXTorr D300-5]. The entire system is mounted on a translation
stage that can be retracted for device exchange without disturbing the optical paths
around the chamber.
The small and moveable vacuum system allows for speed-up of typically extensive procedures
such as device replacement and pumping to UHV, which in our system can take less than one week in total.
The PC is suspended at the tip of a tapered fiber \cite{tiecke_fiber},
which is mounted on a rod assembly and led to the outside of
the vacuum through a fiber feedthrough \cite{cornell_fiber}.
The probe light is sent to the cavity through a beam-sampler
and coupled into the fiber, where the reflection is collected through the
transmission port and onto a single-photon counter [PerkinElmer SPCM-AQR-16-FC].
The reference light for cavity stabilization is combined with the probe on
a dichroic before coupling into the fiber and is monitored in reflection
with a high-bandwidth avalanche photodiode [Menlo Systems APD210].

\section{II. Photonic crystal cavities}
\subsection{Design and fabrication}

\begin{figure*}[htb]
\includegraphics[width=12cm]{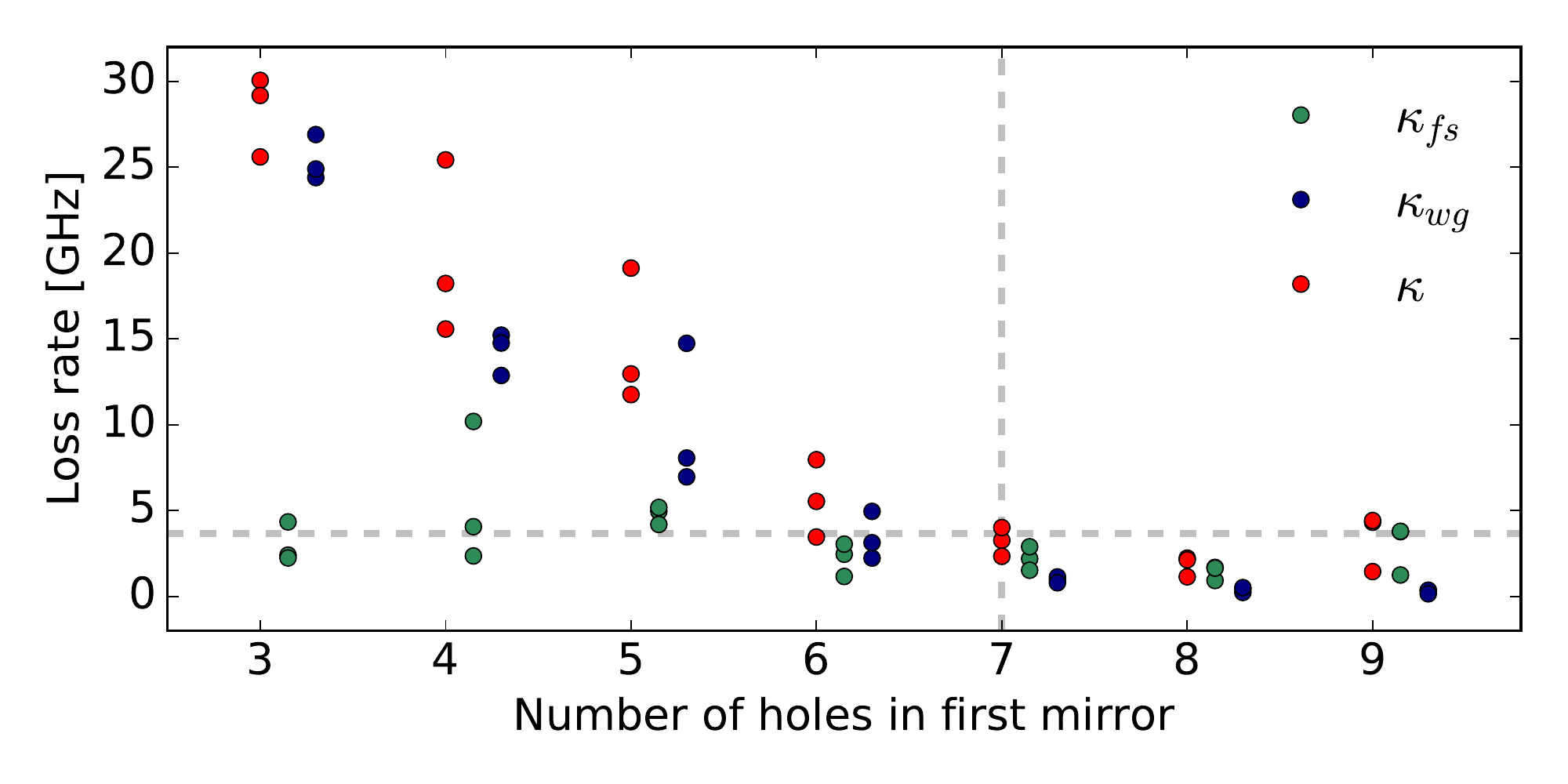}
\caption{Modifying the first mirror loss with the number of holes in the first mirror of the PC and comparing the loss rates for several devices with each hole number.
Grey lines are drawn to indicate the number of holes chosen for the current cavity and its total loss of $\kappa = 2\pi\times$3.68~GHz.}
\label{fig_mirror_tuning}
\end{figure*}

The design of the cavity follows that in \cite{painter_design, painter_zipper}.
The cavity used in this work is designed to operate near the critical coupling regime
($\kappa_{wg} \approx \kappa_{fs}$) for a minimal
total loss rate $\kappa =\kappa_{wg} + \kappa_{fs}$.
Since $\kappa_{fs}$ is limited by fabrication imperfections,
we decrease $\kappa_{wg}$ by controlling the loss from the first Bragg mirror
with the number of holes that define it, as shown in Fig.~\ref{fig_mirror_tuning}.
To reach the critical coupling regime,
$\kappa_{wg}$ is tuned to be comparable to the loss from the cavity into
free space $\kappa_{fs}$ which has typical values of
several gigahertz. The device used
in this work has $\kappa_{wg} = 2\pi\times 0.86$~GHz and $\kappa_{sc}=2\pi\times 2.77$~GHz.

The fabrication process begins with a 200~nm layer of silicon nitride
(bulk index of refraction = 2.02) on a silicon wafer
[Silicon Valley Microelectonics, Inc.].
A pattern of 500 devices is created by exposing a layer of resist
[ZEP520A] on the wafer to a scanning electron beam [Elionix F125].
The pattern is then transferred to the silicon nitride layer using
reactive ion etching.
The silicon underneath is removed in a wet etch process with KOH.
At the end of the process, we deposit an absorptive material (amorphous silicon) onto the disk-shaped heater pad region for thermal tuning of the resonance.

\subsection{Frequency stabilization of the cavity}

\begin{figure*}[htb]
\begin{tikzpicture}
    \draw (0, 0) node[inner sep=0] {\includegraphics[width=\textwidth]{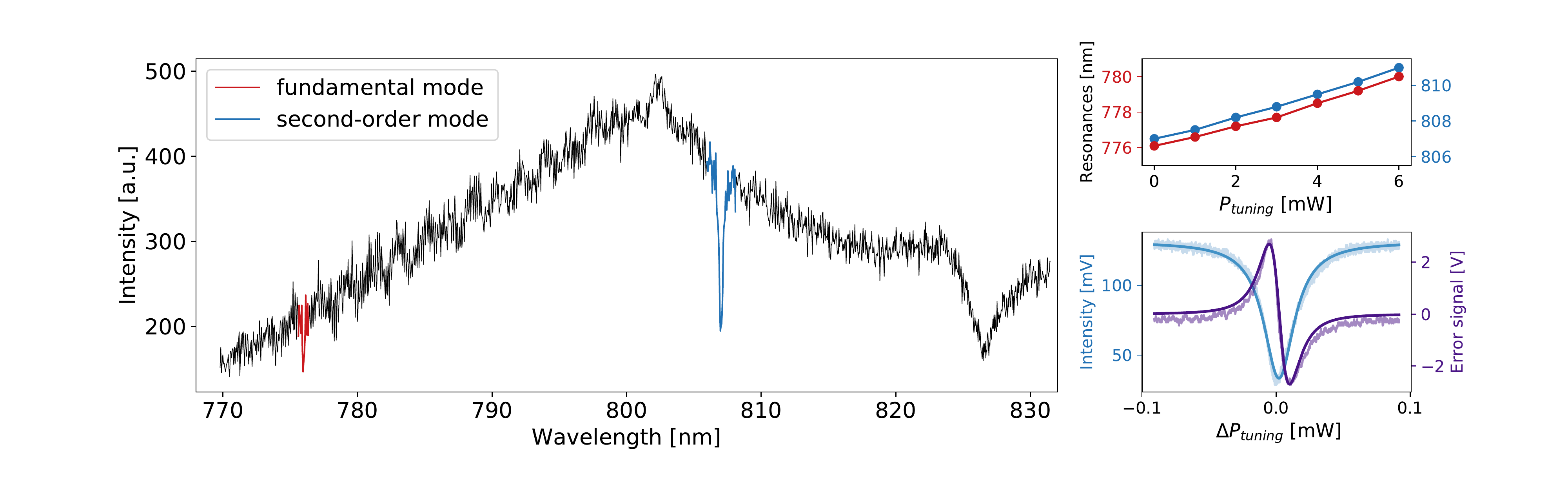}};
    \draw (-7.5, 2.75) node {\normalsize  {\bf a)}};
    \draw (3.45, 2.75) node {\normalsize  {\bf b)}};
    \draw (3.45, 0.25) node {\normalsize  {\bf c)}};
\end{tikzpicture}
\caption{
{\bf a)} Different resonances of the PC cavity observed by sending broadband light and collecting in reflection on a spectrometer. {\bf b)} Simultaneous tuning of the first (red) and second (blue) order resonant wavelengths by increasing the tuning laser power incident on the heater pad.
{\bf c)} A reflection spectrum of the second order resonance (blue) and the PDH signal from the spectrum (purple) obtained from sweeping the tuning laser power over the resonance.}
\label{fig_res_tuning}
\end{figure*}

The resonance of the fundamental mode of the nanophotonic cavity sits near the
$5S_{1/2} \rightarrow 5P_{3/2}$ transition at 780~nm,
but there exist higher-order modes at longer wavelengths as shown in
Fig.~\ref{fig_res_tuning}a.
For the device used in this work, the second-order mode has a resonance near 810~nm,
which is off-resonant for the atoms, and has a linewidth of $2\pi\times$17~GHz.
A laser pointed at the heater pad region heats
up the device and changes the index of refraction, tuning the
the resonances by the same increment, as shown in Fig.~\ref{fig_res_tuning}b.
The bandwidth of the feedback loop is limited by
this thermal tuning process at several milliseconds.

We stabilize the cavity monitoring the second-order mode and
feeding back to the tuning laser power.
The error signal is generated by modulating the phase of the
reference light at 810~nm continuously sent to the cavity at 1~GHz.
The reflected signal from the monitoring avalanche photodiode is
demodulated at the same frequency, creating a low-modulation Pound-Drever-Hall error signal
(Fig.~\ref{fig_res_tuning}c) \cite{pdh_review}.
In the experimental sequence described in the next section,
the cavity resonance must be tuned in and out of resonance
during each trial for different probe pulses.
We achieve this by jumping the setpoint
and waiting for several milliseconds for the cavity to settle before probing.
We have observed no degradation in the quality factor
after inserting our current device within the vacuum system
and continuously tuning and stabilizing its resonance for over one year.

\section{III. Experimental sequence}

\begin{figure*}[htb]
\vspace{15pt}
\includegraphics[width=\textwidth]{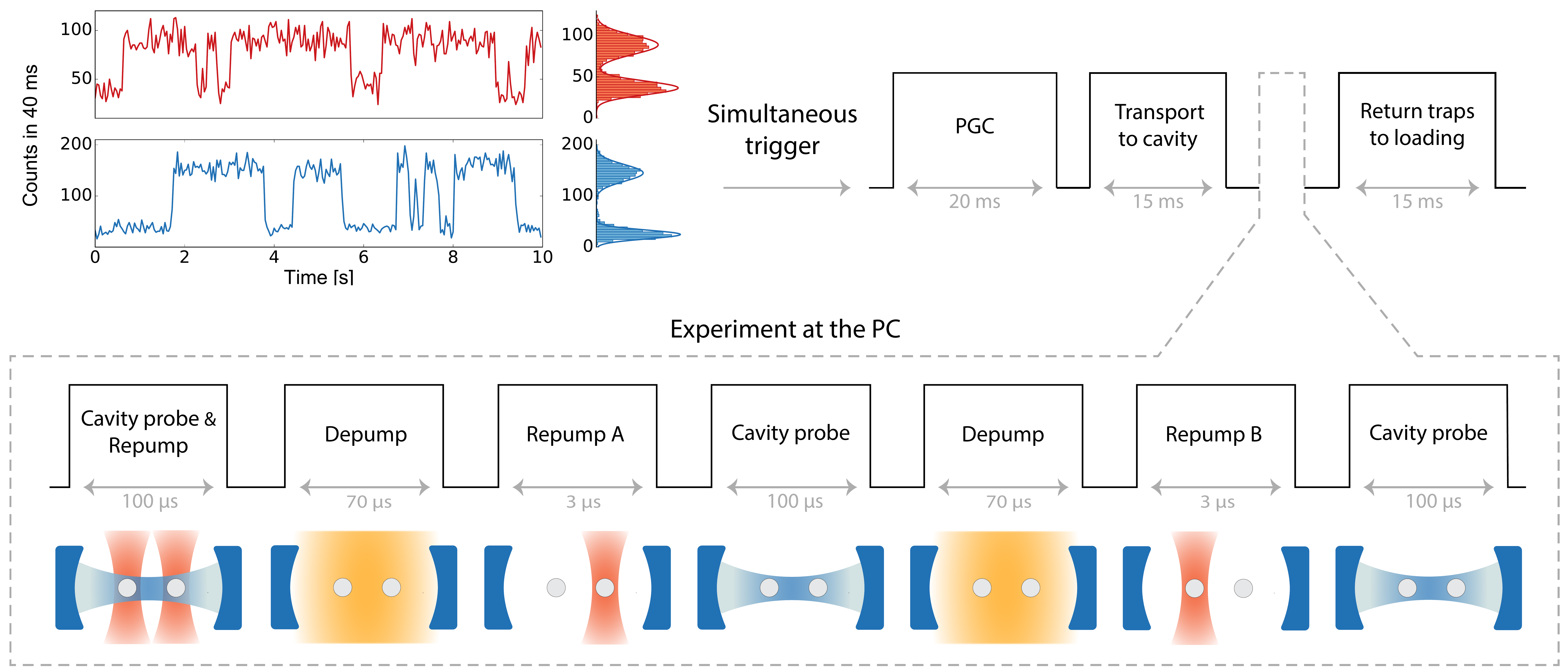}
\caption{
Experimental run is triggered upon rising edges of fluorescence counts from the atoms,
followed by polarization gradient cooling (PGC).
The sequence when the atoms are loaded next to the PC is depicted below,
indicating which beams are global and which are local.
The individual repumping beams are tuned to the $5S_{1/2},~F=1 \rightarrow 5P_{1/2},~F'=2$ transition,
the global depumping beam to the $5S_{1/2},~F=2 \rightarrow 5P_{3/2},~F'=1$ transition.
The data for spectra is taken from the first probe phase,
while the second and third probe phases are used for individual atom post-selection.
}
\label{fig_exp_sequence}
\end{figure*}

We load atoms into the optical tweezers from a magneto-optical trap around the PC.
Operating in the collisional blockade regime \cite{blockade},
we monitor the fluorescence from each of the tweezers with our detector
[Excelitas SPCM-AQRH-14-FC] and distinguish between having zero and one atom
in each tweezer with a fidelity of $>99~\%$.
Every repetition of the experiment begins by triggering on the successful loading of both atoms
(Fig. \ref{fig_exp_sequence}), followed by a 20~ms period of polarization gradient cooling
to bring the atom's temperature down from 50~$\mu$K to 15~$\mu$K.

The atoms are transported to the final positions on the PC by steering
the individual galvanometer mirrors.
Given that our cavity mode is 500~nm wide and 4~$\mu$m long,
accurate positioning of the traps relative to the mode is crucial for the experiment.
The position of the PC on the fiber tip can drift at a rate of $\sim$~100~nm/hour
relative to the microscope objective.
To correct for this drift, we pause the experiment every 20 minutes
and acquire confocal images of the PC at different focal planes.
The images are processed to determine the new position of the PC in three dimensions.
In addition, every few days, we calibrate the atomic coupling strengths
versus position and reposition the atoms such
that they have equal coupling strength.

To take the reflection spectra, we probe the cavity with light resonant with $2 \rightarrow 3'$ transition while constantly depleting the $F=1$ manifold with the individual repumpers.
After acquiring the spectrum, we post-select the data for which an atom was loaded next to the cavity,
which occurs with $\sim$ 70$\%$ probability and is mainly limited by the initial temperature in the traps.
We first prepare both atoms in the uncoupled ground state level $F=1$ with a global beam resonant with the $5S_{1/2},~F=2 \rightarrow 5P_{3/2},~F'=1$ transition.
Then we individually pump a single atom into the coupled ground state $F=2$
with a beam resonant with the $5S_{1/2},~F=1 \rightarrow 5P_{1/2},~F'=2$ transition, co-propagating with the optical tweezers.
After this selective pumping, we probe the cavity on resonance to determine
the atom's presence in a single shot (Fig.~1d, main text).
Repeating this procedure for the other atom allows us to select
for data where one, the other, or both atoms are present.
Once the experiment at the PC is finished, the tweezers are transported back
to the original loading positions. The sequence is outlined in Fig.~\ref{fig_exp_sequence}.

The probe experiment taking place at the cavity typically takes 500~$\mu$s,
much shorter than the lifetime of atoms in the trap next to the PC
of $\sim100$~ms. Compared with the lifetime of 1.5~s away from the PC,
the reduced lifetime next to the PC indicates additional heating mechanism,
as previously observed by \cite{heating_nano}.
Understanding the sources of extra heating and developing cooling techniques
while next to the PC is a subject for future exploration.

\section{IV. Individual tuning of the atomic resonances}
\begin{figure*}[htb]
\begin{tikzpicture}
    \draw (0, 0) node {\includegraphics[width=17.25cm]{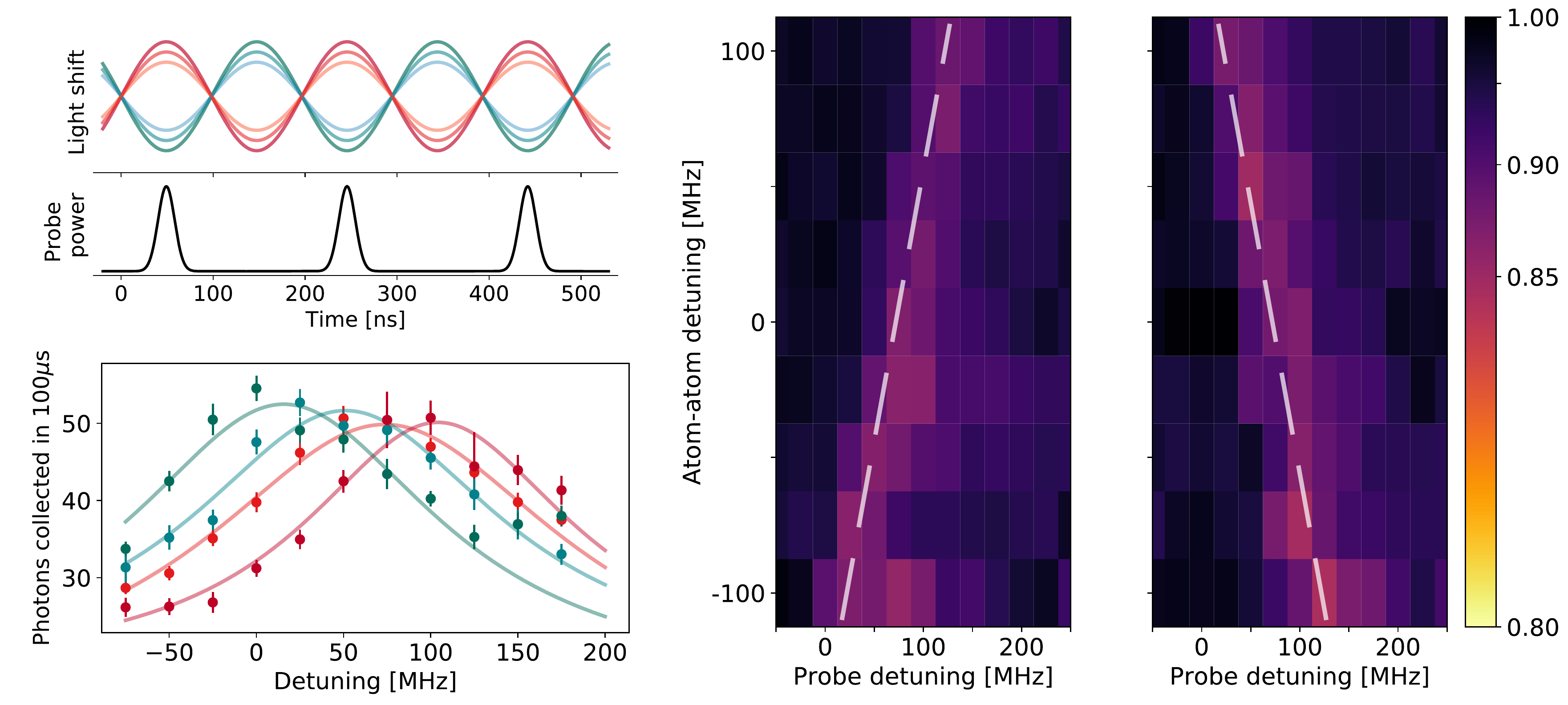}};
    \draw (-8.6, 3.6) node {\normalsize  {\bf a)}};
    \draw (-8.6, 0) node {\normalsize  {\bf b)}};
    \draw (-1.25, 3.6) node {\normalsize  {\bf c)}};
\end{tikzpicture}
\vspace{-15pt}
\caption{
{\bf a)} Illustration of the instantaneous light shift experienced by each atom probed while modulating the individual tweezer intensity.
{\bf b)} Reflection spectra in the resonant regime at two different modulation contrasts (light/dark) for one atom (greens) and the other (reds) to show how the individual light shifting can be tuned simultaneously.
{\bf c)} Experimental data of reflectivity map over probe detuning and
atom detuning in the dispersive regime showing light shift tuning.
Fig.4a in the main text
is generated from the average of these two plots.}
\label{fig_trap_mod}
\end{figure*}

The independent optical tweezer paths in our setup allow
for individual manipulation of the internal atomic states.
In this work, we change the relative atomic detuning $\delta_{12}$
through light shifts from the individual tweezers. This is done by modulating their
intensities out of phase. Tuning the contrast of the modulation then shifts the atoms
in and out of resonance with respect to each other.
Note that we choose the modulation frequency
$\Delta_{\text{mod}} \approx 2\pi\times5~\text{MHz}$, which is faster than the motional frequencies of the traps,
thereby subjecting the two atoms to the same average trapping potential.
The intensity modulation is generated using a Mach-Zehnder interferometer,
with an acousto-optical modulator (AOM) driven at
$\omega_{\text{AOM}}\pm\Delta_{\text{mod}}/2$ in each arm, where
$\omega_{\text{AOM}}~=~2\pi\times80~\text{MHz}$ is the central frequency of the AOMs.
The relative phase between the two tweezers can be tuned by sending
one of the outputs of the interferometer through a delay line ($\sim100~\text{m}$)
and fine tuning the relative drive frequency $\Delta_{\text{mod}}$
to match the path difference.
The modulated optical signal is also used to trigger the probe light sent to the cavity.
Calibrating the delay of the triggered probe pulse allows us to probe the atoms at their maximum/minimum light shifts.
When operating with the traps out of phase, changing the contrast of this modulation changes the individual lightshift of each atom and is used to sweep the atom-atom detuning as depicted in Fig. \ref{fig_trap_mod}.

\section{V. Theoretical models}

To model the reflection spectra, we study the transitions between the
$5S_{1/2},~F=2$ ground state manifold, labeled $\ket{g}$, and
the relevant excited states in $5P_{3/2}$ i.e. $F'=1,2,3$,
labeled $\ket{e_i}$, where $i = 1,2,3$.
For the on-resonance measurements (FIGs. 1 and 2 in the main text),
the cavity frequency $\omega_c$ is resonant with the $2 \rightarrow 3'$ transition.
The cavity field decay rate ($\kappa = 2\pi\times3.7$~GHz)
is smaller than the hyperfine splitting between the $F=1$ and $F=2$
($\Delta_{HF} = 2\pi\times6.8 $~GHz), so we neglect the cavity coupling to $F=1$ manifold.
Experimentally, we confirm this by noting that when the atom is in $F=1$,
the cavity reflectivity is the same as if there were no atom coupled to it.

A system consisting of a cavity and a multi-level atom is described by the Hamiltonian \cite{kimble_hyperfine}:
\begin{equation}\label{H}
H = \omega_c a^{\dagger}a + \sum_{i} \omega_i \sigma_i^{\dagger}\sigma_i + \sum_i g_i(a^{\dagger}\sigma_i + a \sigma_i^{\dagger})
\end{equation}
where $\sigma_i  =\ket{g}\bra{e_i}$ and $a$ are the lowering operators
for the atomic and cavity excitations respectively,
$g_i$ is the single-photon Rabi frequency between the cavity and the excited state $\ket{e_i}$, and $\omega_i$ is the resonance frequency between $\ket{g}$ and $\ket{e_i}$. Each excited state decays into free space at a rate $\gamma$, and cavity decays at a total rate $\kappa$, out of which $\kappa_{wg}$ is into the collected waveguide mode.

To model the incoherent population decays from the atomic excited states and the cavity mode,
we define the Lindblad jump operators $ L_0 = \sqrt{\kappa}a$ and $L_i = \sqrt{\gamma}\sigma_i$.
We use individual decay operators $L_i$ instead of the cumulative decay operator $\sum_i L_i$ because
$\gamma = 2\pi\times6 $~MHz is much smaller than the hyperfine splittings in the excited states of
$2\pi\times\{267,~157\}$~MHz, so we can neglect the interference between the spontaneously emitted photons.
We confirm this by numerically modelling and comparing both cases.

The system dynamics can be described with Heisenberg-Langevin equations of the form
\begin{equation}\label{heis-lan}
    \dot{A} = i [H,A]+
    \sum_{i} \Big(L_i^{\dagger}A L_i -\frac{1}{2}(A L_i^{\dagger}L_i + L_i^{\dagger}L_i A)\Big)
\end{equation}
where $A \in \{a,\sigma_i\}$.
After solving for $a$, the reflectivity can be obtained from the input-output relation:
\begin{equation}\label{inout}
    a_{out}+a_{in} = \sqrt{\kappa_{wg}} a
\end{equation}
\begin{equation}\label{reflectivity}
    r = \frac{a_{out}}{a_{in}} = \sqrt{\kappa_{wg}}\frac{a}{a_{in}} - 1
\end{equation}

The resulting system of equations reads:
\begin{equation}\label{dot_a}
    \dot{a} = -i \omega_c a -i \sum_{j}g_j\sigma_j -\frac{\kappa}{2}a + \sqrt{\kappa_{wg}}a_{in}
\end{equation}
\begin{equation}\label{sig_i_full}
    \dot{\sigma_i} = - i \omega_i\sigma_i +
    i a \sum_j g_j \Big(\ket{e_j}\bra{e_i} - \delta_{i,j}\ket{g}\bra{g}\Big) - \frac{\gamma}{2}\sigma_i
\end{equation}
We are interested in the frequency response of the reflection
to a weak drive of frequency $\omega$. Since the steady-state population in $\ket{e_i}$ is negligible
in this weak excitation limit, we can substitute $\ket{e_j}\bra{e_i} - \delta_{i,j}\ket{g}\bra{g} \approx -1$. Solving these equations, we obtain the reflectivity
\begin{equation}
r = \kappa_{wg}
  \Bigg(
  {\frac{\kappa}{2} - i\delta_c + \sum_{i} \frac{g_i^2}{\gamma/2-i\delta_i}}
  \Bigg)^{-1}
  -1,
\label{reflectivity_final}
\end{equation}
where $\delta_c = \omega - \omega_c$ and $\delta_i = \omega - \omega_i$.
The expression has the same format as the one for multiple emitters coupled to the same cavity mode \cite{ruffin}
and can be extended by modifying the sum to include multiple emitters.
The cooperativity associated with the $i$-th transition
is $C_i = 4g_i^2/\kappa\gamma$. The reflection spectra are measured as $|r(\omega)|^2$.

\subsection{Atomic motion in the cavity mode}
Nanophotonic structures confine photons to sub-wavelength mode volumes.
The longitudinal confinement is a $\sim 4~\mu$m long gaussian envelope,
and the transverse confinement is given by the evanescent field of decay with
the characteristic length of $z_0 = 120$~nm \cite{jeff_science}.
Moreover, the mode intensity is longitudinally modulated with the lattice constant of $a=290$ nm.
Due to these spatial variations, an atom with thermal motion will be subjected
to a fluctuating coupling strength.
\begin{figure*}[htb]
\begin{tikzpicture}
    \draw (0, 0) node[inner sep=0] {\includegraphics[width=\textwidth]{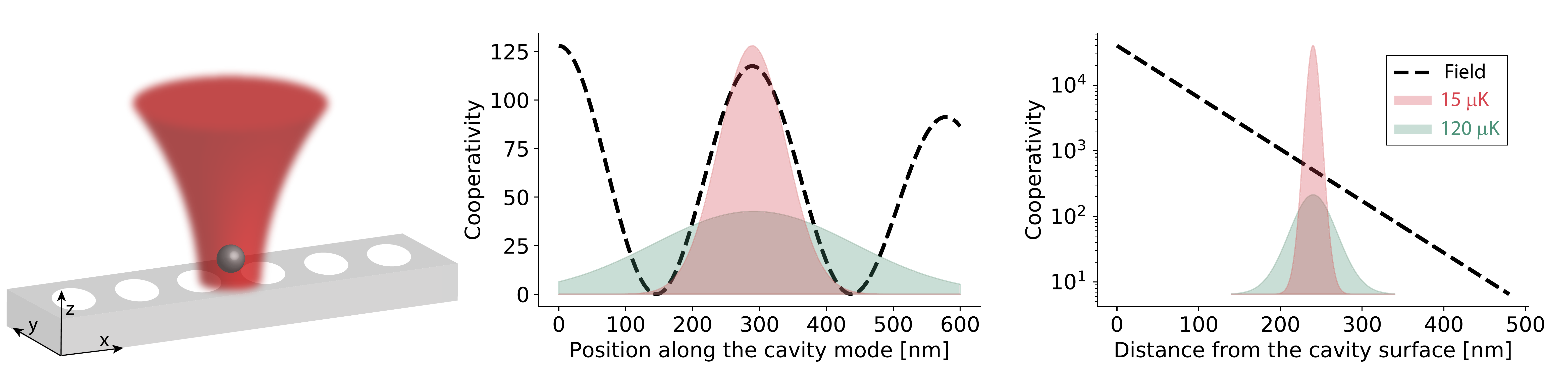}};
    \draw (-8.75, 1.7) node {\normalsize  {\bf a)}};
    \draw (-3.8, 1.7) node {\normalsize  {\bf b)}};
    \draw (2.5, 1.7) node {\normalsize  {\bf c)}};
\end{tikzpicture}
\caption{
{\bf a)} Illustration of an atom trapped in a tweezer above the photonic crystal and definition of axes.
{\bf b)} Cooperativity variation in the longitudinal (x-axis) direction. The green and red dashed lines correspond to the atomic wavefunction size at $120~\mu $K and $15~\mu $K, respectively.
{\bf c)} Cooperativity variation and atomic wavefunction sizes in the $z$ direction. }
\label{fig3:mode_profile}
\end{figure*}

The variation of the atomic position depends on its temperature and confinement length scale.
Given the motional radial frequency measured to be $2\pi\times 115~$kHz,
we obtain the size of of the atomic wavefunction of $50$~nm in the radial direction of the tweezer
at $15~\mu$K (our temperature after polarization gradient cooling).
As the atoms are transported to the
cavity, the trapping potential gains additional
axial confinement from the reflection of the tweezer
off of the structure \cite{jeff_science}.
The process of potential distortion from a gaussian beam
to a standing wave can introduce heating to the atom.
From separate measurements of the coherence times
between two magnetic sublevels in the ground state
at varying tweezer depths similar to \cite{temp_est}, we estimate the upper bound
for our temperature at the cavity to be $120~\mu$K,
corresponding to a wavefunction size of $150$~nm in the radial direction of the tweezer.

The standing wave formed in the axial direction results in a higher axial frequency
of $2\pi\times 550~$kHz, and a spatial confinement of $195$~nm.
This tight confinement is crucial for the operation
since the evanescent field decays sharply in the axial direction.
Fig.~\ref{fig3:mode_profile} shows the variation of the cavity mode in the x- and z- directions,
overlapped with our estimates of the lower and upper bound on wavefunction size at $15~\mu$K and $120~\mu$K.

To account for the cavity field sampling, we model the cooperativity $C$, which is proportional to the field intensity $I$, to vary in the two directions as
\begin{equation}
    C(x,z) = C_0 \cos^2{\Big(\frac{\pi x}{a}\Big)}e^{-2z/z_0}
    \label{C_dependence}
\end{equation}
where $a = 290$~nm, $z_0 = 120$~nm, and $C_0$ is the cooperativity the atom would experience
in the absence of motion at its center position $(x,z)=(0,0)$.
Note that we have neglected the slowly-varing
envelope in the direction along the standing wave with the characteristic mode length
of 4~$\mu$m as well as the variation in the $y$ direction.

\subsection{Cooperativity estimation}

To model our experimental reflection spectrum (e.g. Fig. 1d, main text)
and extract an estimate for the associated cooperativity, we
combine the expression for reflectivity in Eq.(\ref{reflectivity_final})
with the spatial dependence of the cooperativity in
Eq.(\ref{C_dependence}).
To produce a spectrum, we sample atomic positions in the $x$ and $z$ directions
(as defined in Fig. \ref{fig3:mode_profile}a) from normal distributions with standard deviations
$w_x$ and $w_z$ respectively and generate a reflectivity averaged over multiple values of
the resulting cooperativities.
The fit parameters consist of $w_x$, $w_z$, and the cooperativity amplitude
$C_{0, i}$ in Eq.(\ref{C_dependence})
associated with transition to each excited state $\ket{e_i}$.

\begin{figure*}[htb]
\begin{tikzpicture}
    \draw (0, 0) node[inner sep=0] {\includegraphics[width=\textwidth]{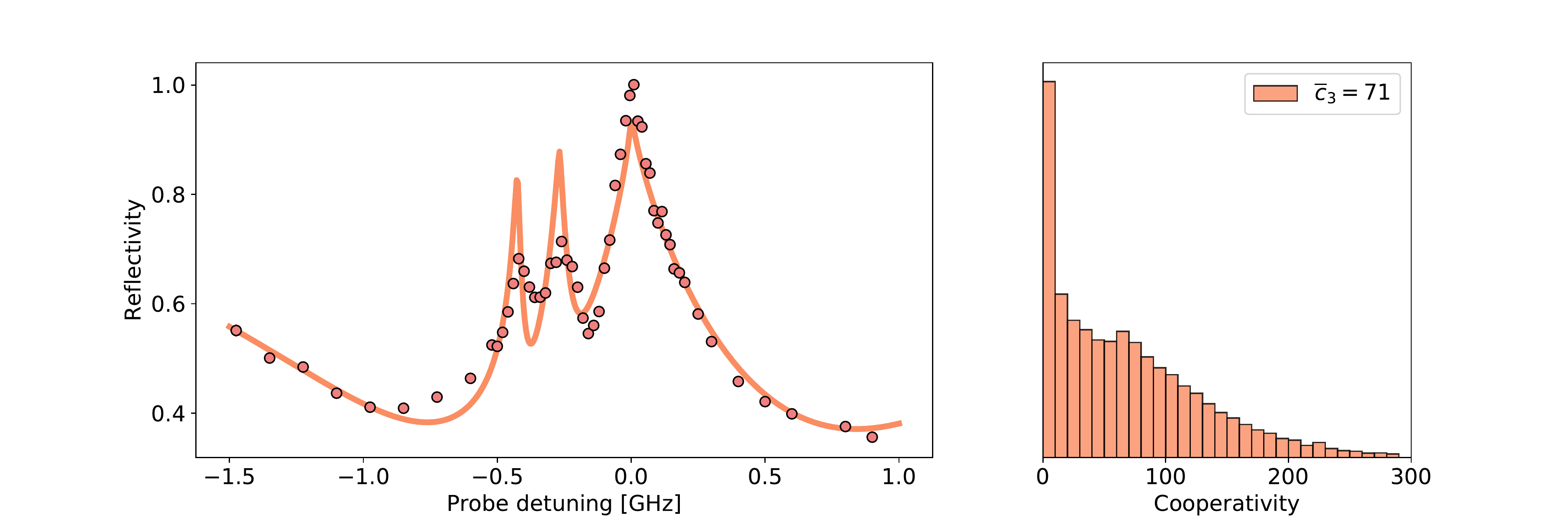}};
    \draw (-7.8, 2.1) node {\normalsize  {\bf a)}};
    \draw (2.4, 2.1) node {\normalsize  {\bf b)}};
\end{tikzpicture}
\caption{{\bf a)} Theoretical model for the single-atom reflection spectrum at the mode maximum overlapped with experimental data. {\bf b)}
Associated cooperativity distribution with the mean of 71(4) and
standard deviation of 66(4).}
\label{fig1:refl_spec}
\end{figure*}

The best fit (Fig. \ref{fig1:refl_spec}a) for the data according to this model
yields $w_x = 190$~nm, $w_z = 33$~nm, consistent with our temperature estimates of $120~\mu$K.
The cooperativity associated with the $3'$ line of interest
is $C_{0,i=3} = 128(6)$ in the absence of motion.
The histogram of the cooperativity distribution
associated with the spectrum is shown in Fig. \ref{fig1:refl_spec}b,
yielding the mean of $\langle C_{i=3} \rangle = 71(4)$,
and the standard deviation of 66(4).
The uncertainties are determined by bootstrapping.

The value of $C_{0}$ extracted is consistent with our
independent estimate of the field strength at the surface of the PC of
$2g_\text{max} \approx 2\pi\times15~\text{GHz}$
which evanescently decays down to
$2g_{0} \approx 2\pi\times1.7~\text{GHz}$ at 260~nm away from the PC.
The single-photon Rabi frequency at the surface $2g_\text{max}$
(corresponding to the maximum cooperativity of $C_\text{max} \approx 40,000$)
is calculated from the single-photon electric field of
$E_\text{max} \approx 2~\text{kV/cm}$, estimated for the mode volume
$V_\text{mode} = 0.4 \lambda^3$ \cite{jeff_science}.
The biggest deviation from the model occurs in the height of the $1'$ and $2'$ lines.
Some of the effects neglected in this model include decays to $F=1$ ground state
and repumping from it (which could explain the reduced heights),
the motion in the $y$ direction, heating during probing, and trap pointing uncertainty.

\begin{figure*}[h]
\begin{tikzpicture}
    \draw (0, 0) node[inner sep=0] {\includegraphics[width=\textwidth]{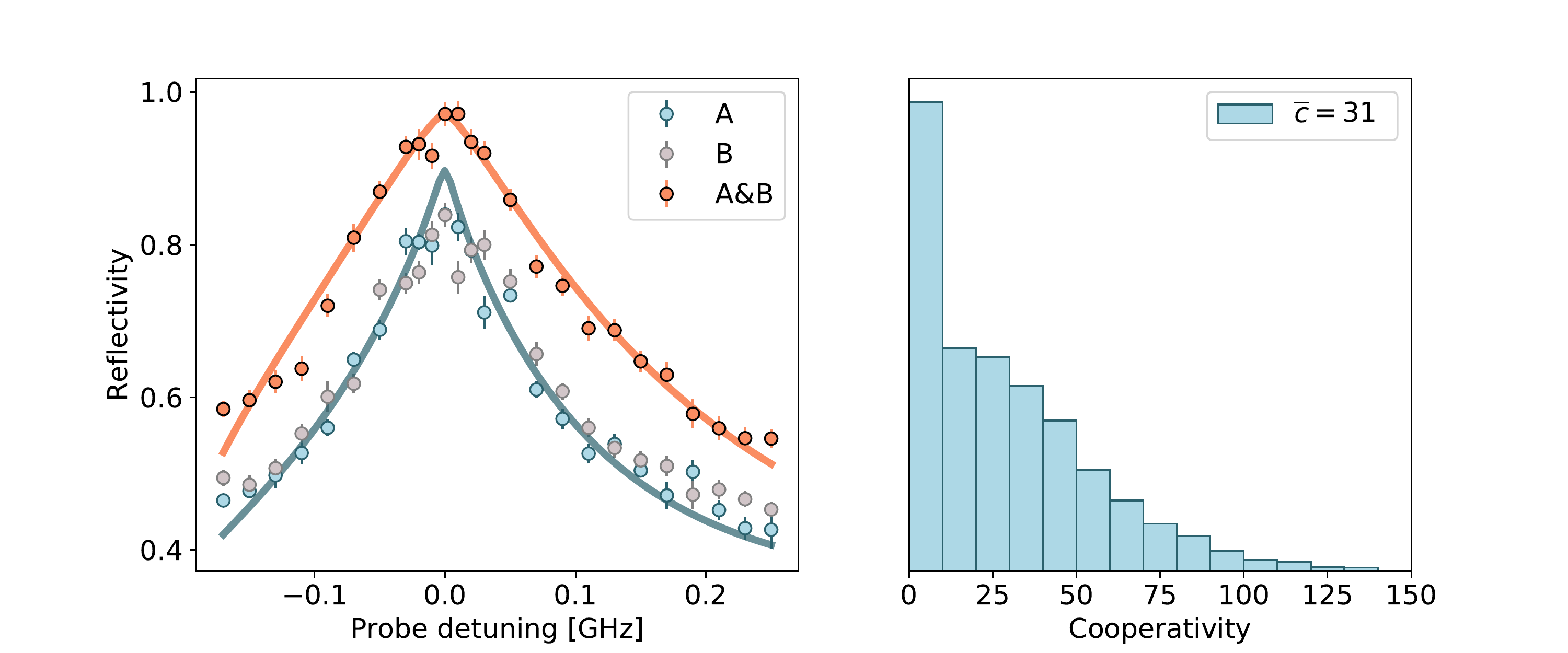}};
    \draw (-7.9, 2.6) node {\normalsize  {\bf a)}};
    \draw (0.9, 2.6) node {\normalsize  {\bf b)}};
\end{tikzpicture}
\caption{{\bf a)} Theoretical model for the single-atom reflection spectrum $1~\mu$m away from the mode center (blue) and for the two-atom reflection spectrum (red) generated using the same cooperativity distribution. {\bf b)} Distribution of single-atom cooperativity for the $2\rightarrow 3'$ transition.}
\label{fig2:superradiance}
\end{figure*}

The two-atom spectrum (FIG 2, main text) is taken with
atoms placed $1~\mu$m away from the mode center.
We now focus on the $3'$ line, which is also the transition we use in the dispersive regime.
The fit to the single-atom reflectivity at this position is shown in the
Fig. \ref{fig2:superradiance}.
The wavefunction size $w_x = 190$~nm, $w_z = 33$~nm,
is the same as in the Fig.~\ref{fig1:refl_spec}, whereas $C_{0,i=3} = 56(3)$,
corresponding to the mean single-atom cooperativity of
$\langle C_{i=3} \rangle  = 31(2)$
and the standard deviation of $28(2)$, averaged over two atoms.
The reduction of the mean cooperativity is expected from the mode profile shown in Fig. 2c in the main text.
The two-atom reflection spectrum is generated based on
both single-atom spectra with no additional parameters.

The spectra in the dispersive regime taken with nonzero light shifts
(Fig. 3b, 4, main text) are analyzed with the same model.
In addition, we take into account the effect of fluctuating light shift $\delta_{A,B}$
due to atomic motion.
We extract $J = 2\pi\times25(4)$~MHz at $\Delta = 2\kappa$,
consistent with our cooperativity estimate for the resonant regime.

\clearfmfn


\begin{thebibliography}{47}
\bibitem [{Note1()}]{0mhzshift}%
  \BibitemOpen
  \bibinfo {note} {The shift from 0~MHz detuning is observed in the absence of
  the light shift from the tweezer.}\BibitemShut {Stop}%
\bibitem [{Note2()}]{50mhzshift}%
  \BibitemOpen
  \bibinfo {note} {The shifts are determined with respect to an offset of $2\pi
  \times 50~\protect \text {MHz}$ due to the light shift induced by the
  tweezer.}\BibitemShut {Stop}%
\end{thebibliography}
\end{document}